# Entwined lattice of atoms and anionic electrons in layered electride LaCl

Songyuan Geng[†1], Xin Wang[†1], Jianqi Zhong[†1], Risi Guo[1], Fangjie Chen[1], Qun Wang[1], Kangjie Li[1], Keyu An[1], Teng-Fei Ying[1], Chen Qiu[2], Hanpu Liang[3], Zhengtai Liu[4,5], Mao Ye[4,5], Sungsoo Hahn[6], Balasubramanian Thiagarajan[6], Benjamin T. Zhou[1] and Haoxiang Li[*1]

[1]Advanced Materials Thrust, Function Hub, The Hong Kong University of Science and Technology (Guangzhou), Guangzhou 511453, China

[2]Department of Physics, Eastern Institute of Technology, Ningbo 315200, China

[3]Advanced Interdisciplinary Science Research Center, Ningbo Institute of Materials Technology and Engineering, Chinese Academy of Sciences, Ningbo 315201, China

[4]Shanghai Synchrotron Radiation Facility, Shanghai Advanced Research Institute, Chinese Academy of Sciences, Shanghai 201204, China

[5]National Key Laboratory of Materials for Integrated Circuits, Shanghai Institute of Microsystem and Information Technology, Chinese Academy of Sciences, Shanghai 200050, China

[6]MAX IV Laboratory, Lund University, Lund SE-221 00, Sweden

[†]These authors contributed equally to this work.

[*]Corresponding authors. Email:

haoxiangli@hkust-gz.edu.cn

**ABSTRACT: Controlling the lattice geometry that governs electronic structure is a central theme in condensed-matter physics, yet in crystalline solids this geometry is usually fixed by the atomic framework. Electrides offer an alternative route to electronic structure design in which their excess electrons can organize into anionic electron lattice (AEL) and provide a lattice-like degree of freedom. Recent work has highlighted the "standalone" limit, where the AEL in YCl yields bands well described by the dice-lattice model. Here, using angle-resolved photoemission spectroscopy (ARPES), we show that LaCl, although isostructural to YCl, realizes a qualitatively different regime where the AEL is entwined with the La cation framework, producing a fully reconstructed electronic structure. Combining the ARPES result with tight-binding model analysis, we demonstrate that this radical divergence stems from the activation of direct hopping channels between the AEL and the La atomic lattice. This coupling reshapes the effective lattice geometry, reconstructs the electronic states, and modifies the associated Chern band topology, transforming the bipartite dice-lattice network in YCl into a tripartite structure in LaCl. Our findings demonstrate that the coupling between the AEL and the atomic lattice can actively shape the effective lattice geometry that governs the electronic structure. This coupling can act as a powerful tuning knob for electronic structure design that is inaccessible in conventional materials.**

In crystalline solids, lattice geometry has long served as a guiding principle for understanding and engineering electronic structure [1-6]. However, this geometry is usually fixed by the atomic framework [7, 8]. Once the atomic framework is fixed, the resulting electronic structure is expected to follow the site connectivity and network topology [9-15]. Chemical substitutions, orbital hybridizations and lattice distortions may serve as tuning knobs, but the qualitative architecture of the electronic bands is generally constrained by geometry [16-22]. This geometry-centered paradigm therefore enables the targeted exploration of unusual electronic states driven primarily by lattice geometry, such as in the kagome materials and moiré materials [23-29].

This conventional correspondence between atomic geometry and electronic structure breaks down in electrides, where excess valence electrons localize in lattice voids as pseudo-anions (Fig. 1a) [30-36]. In layered electrides, these interstitial anionic electrons (IAEs) can organize into an anionic electron lattice (AEL) that form a geometrical structure distinct from the atomic lattice and dominate the low-energy electronic states [37-47]. A striking realization of this behavior is the van der Waals (vdW) electride $[YCl]^{2+}\cdot 2e^{-}$ (YCl) [48]. In YCl, although the atomic backbone forms a honeycomb-like lattice (see supplementary information Fig. S1), the IAEs self-assemble into a standalone AEL of dice-lattice geometry (Fig. 1b). Correspondingly, the low-energy electronic structure exhibits the characteristic dice-lattice bands, including the dice-lattice flat bands near the Fermi level (Fig. 1b, c) [48, 49]. YCl therefore establishes the standalone-AEL limit, in which the low-energy electronic structure is governed primarily by the AEL rather than by the atomic lattice.

Traditionally, once the lattice geometry is fixed, the effective lattice connectivity is determined, and the electronic structure is expected to follow [50]. As the $[REX]^{2+}\cdot 2e^{-}$ family of electrides (REX, RE = rare-earth metal, X = halogen, Fig, 1a) shares the same crystallographic symmetry and symmetry-related interstitial sites as YCl [51-54], one would naturally expect similar dice-lattice-derived bands across the family, with the standalone AEL serving as the effective geometric constraint on the electronic structure. Nevertheless, $[LaCl]^{2+}\cdot 2e^{-}$ (LaCl) defies this expectation. Despite being isostructural and isoelectronic to YCl, LaCl exhibits a reconstructed electronic structure (Fig, 1f) and altered Chern band topology (Fig, 1d,g), indicating that in layered electrides, the AEL is highly tunable, and the electronic structure cannot be easily inferred from the standalone electron lattice.

In this work, we show that despite sharing the same atomic framework and IAE arrangement as YCl, all low-energy bands in LaCl depart from the YCl dice-lattice electronic structure, as revealed by our ARPES measurements (Fig. 1b,f). These two compounds present a striking contrast, sharing identical lattice symmetry and electron filling while exhibiting fully distinct electronic structures near $E_{\mathrm{F}}$. Tight-binding (TB) modeling further reveals that the reconstructed band structure of LaCl originates from direct hopping pathways between the La sites and the IAE states (Fig. 1e), which are absent in YCl. Consequently, we demonstrate that LaCl realizes an entwined lattice of both atoms and anionic electrons, which fundamentally alters the effective lattice of the crystal, rewires the lattice connectivity and reshapes the electronic structure. As a direct manifestation of this reconfigured network, in LaCl, highly dispersive bands replace the dice flat band structure, and a van Hove singularity emerges at the M point near the Fermi level, and the associated Chern topology is also reconstructed from the original $|C| = 4$ in YCl [49] to $|C| = 3$ in LaCl. This transition from the standalone AEL in YCl to the entwined lattice of LaCl occurs without altering the crystallographic framework, demonstrating that the AEL can serve as an intrinsic structural degree of freedom for engineering electronic structures. More broadly, just as moiré geometry established a powerful design principle for reshaping electronic states in twisted bilayer systems [27, 55-58], the AEL provides an intrinsic and symmetry-compatible route to electronic structure engineering in electrides, enabling a form of lattice engineering that is unavailable in conventional crystals.

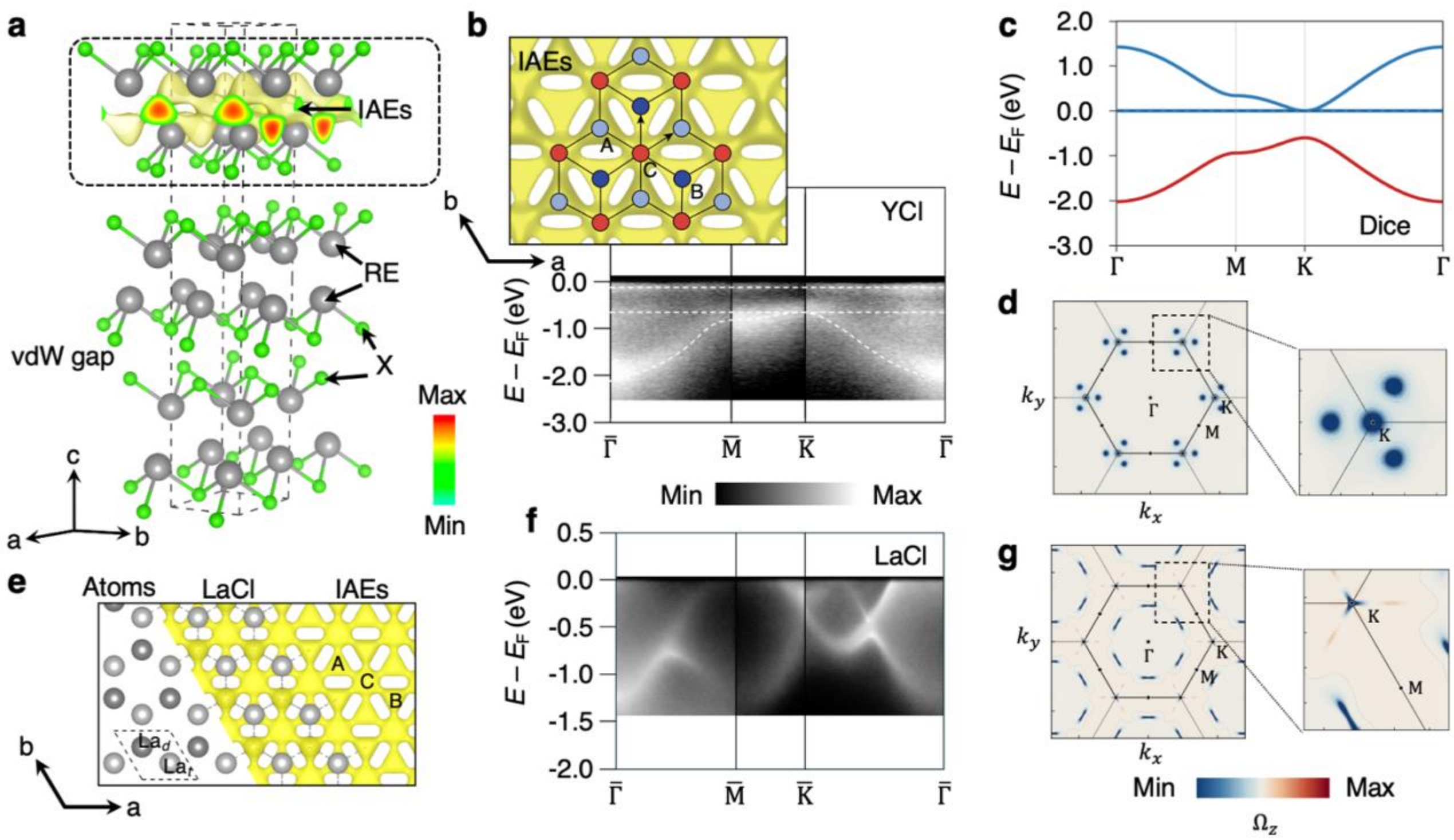

**Fig. 1 From standalone anionic electron lattice in YCl to entwined atom and anionic electron lattices in LaCl. a.** Crystal structure of REX (RE = rare-earth metal and X = halogen) van der Waals (vdW) electride. The monolayer unit is marked within the dashed box. **b.** The "standalone" anionic electron lattice (AEL) exemplified by YCl, where the interstitial anionic electrons (IAEs) form an effective dice-lattice structure that is distinct from the honeycomb atomic framework. The lower-right inset shows the characteristic dice-lattice flat band near the Fermi level observed by angle-resolved photoemission spectroscopy (ARPES). The dashed-white lines serve as the guide-to-the-eye lines for the dice-lattice bands. **c.** Tight-binding (TB) band structure of the dice lattice, reproducing the main features of the ARPES result. **d.** Calculated Berry curvature distribution of the dice-lattice bands in YCl. The Berry curvature is concentrated near the avoided crossings around the K points, giving rise to the Chern-number-four topology of the standalone AEL state. **e.** Structural and electronic connectivity of the entwined lattice in LaCl. Top-view comparison showing that LaCl retains the same honeycomb cation framework as YCl, while the IAEs occupy the same symmetry-related interstitial sites, as revealed by DFT ELF mapping. Unlike the standalone AEL in YCl, however, LaCl forms an entwined lattice in which the cations and interstitial anionic-electron states are electronically coupled. The side view highlights the additional hopping channels between the La sites and the IAE sites A, B and C, shown by red dashed lines. **f.** ARPES band structure of LaCl. In contrast to the dice-lattice electronic structure of YCl, the low-energy bands in LaCl are strongly reconstructed, demonstrating the breakdown of a standalone AEL description. **g.** Calculated Berry curvature distribution of the reconstructed bands in LaCl. Compared with YCl, the redistribution of Berry curvature reflects the reconstruction of the effective lattice connectivity and the associated change in band topology, yielding a Chern-number-three state. Blue and red denote negative and positive Berry curvature, respectively. Both ARPES measurements were taken using 75 eV photons at low temperature (T < 10 K).

**ARPES evidence for a reconstructed band structure in LaCl**

To reveal the electronic structure of LaCl, Fig. 2 presents a comprehensive analysis of the ARPES data. In LaCl, three dispersive band manifolds labelled $\alpha$, $\beta$, and $\gamma$ dominate the low-energy electronic structure. Along $\bar{\Gamma}$–$\bar{\mathrm{K}}$ direction (Fig. 2a,e,j), the $\alpha$ band crosses $E_{\mathrm{F}}$, forming a large $\bar{\Gamma}$-centered hole pocket (Fig. 2l,m), and turns back toward $E_{\mathrm{F}}$, generating a small electron pocket $\alpha^*$ around the zone corner. The $\beta$ band, forming a clear spectral feature in the $\bar{\Gamma}$–$\bar{\mathrm{M}}$ spectrum, disperses to high binding energy and merges with the $\gamma$ band. The $\gamma$ manifold displays a pronounced non-monotonic dispersion with the spectral features clearly resolved in all three high-symmetry cuts (Fig. 2a-c). The three dominant band dispersions are tracked in the second-derivative maps (Fig. 2e,g,i) and further emphasized in the stacked energy distribution curves (EDCs) (Fig. 2j,k). The measured dispersions in LaCl closely follow the DFT bands (Fig. 2d,f,h). However, the exchange splitting predicted by DFT is not clearly resolved in ARPES, indicating that the intrinsic spin splitting is smaller than the DFT prediction and that any residual splitting in the spectrum is further masked by matrix-element-induced intensity imbalance between the spin-split branches and finite linewidth broadening (see Supplementary Information S5 for further discussion).

In sharp contrast to the dice-lattice electride YCl, in which a flat band spans the entire Brillouin zone at $E_\mathrm{F}$ [48], LaCl exhibits no signatures of the dice bands. Instead, the spectra are dominated by dispersive states along all high-symmetry directions (Fig. 2a-c), demonstrating that LaCl realizes a qualitatively distinct low-energy electronic structure rather than a perturbative variant of the dice bands.

The qualitative difference of the electronic structures in these two compounds does not arise from the variation of electron count or crystallographic symmetry. Single-crystal X-ray diffraction (XRD) and X-ray photoemission spectroscopy (XPS) confirm that LaCl and YCl share similar crystal structures and the nominally trivalent cation state [59], respectively (see Supplementary Information S2 and S3). Furthermore, electron localization function (ELF) of LaCl (see Fig. 1e,g and more discussion in Fig. 4) from DFT reveals excess charge density localized in the same symmetry-related crystallographic voids as in YCl, although their spatial distributions differ because of enhanced coupling to the neighboring La sites (Fig. 4c,d). The low-energy band structure of LaCl (Fig. 3a) is well separated from the Cl-derived states (Fig. 3b), just as in YCl, ruling out any contamination from the chlorine valence bands. Together, these observations indicate that the qualitative difference in the electronic structures of the two compounds does not originate from conventional structural or filling effects, but instead arises from a reconstruction of the effective lattice through additional interactions between the atomic framework and the AEL.

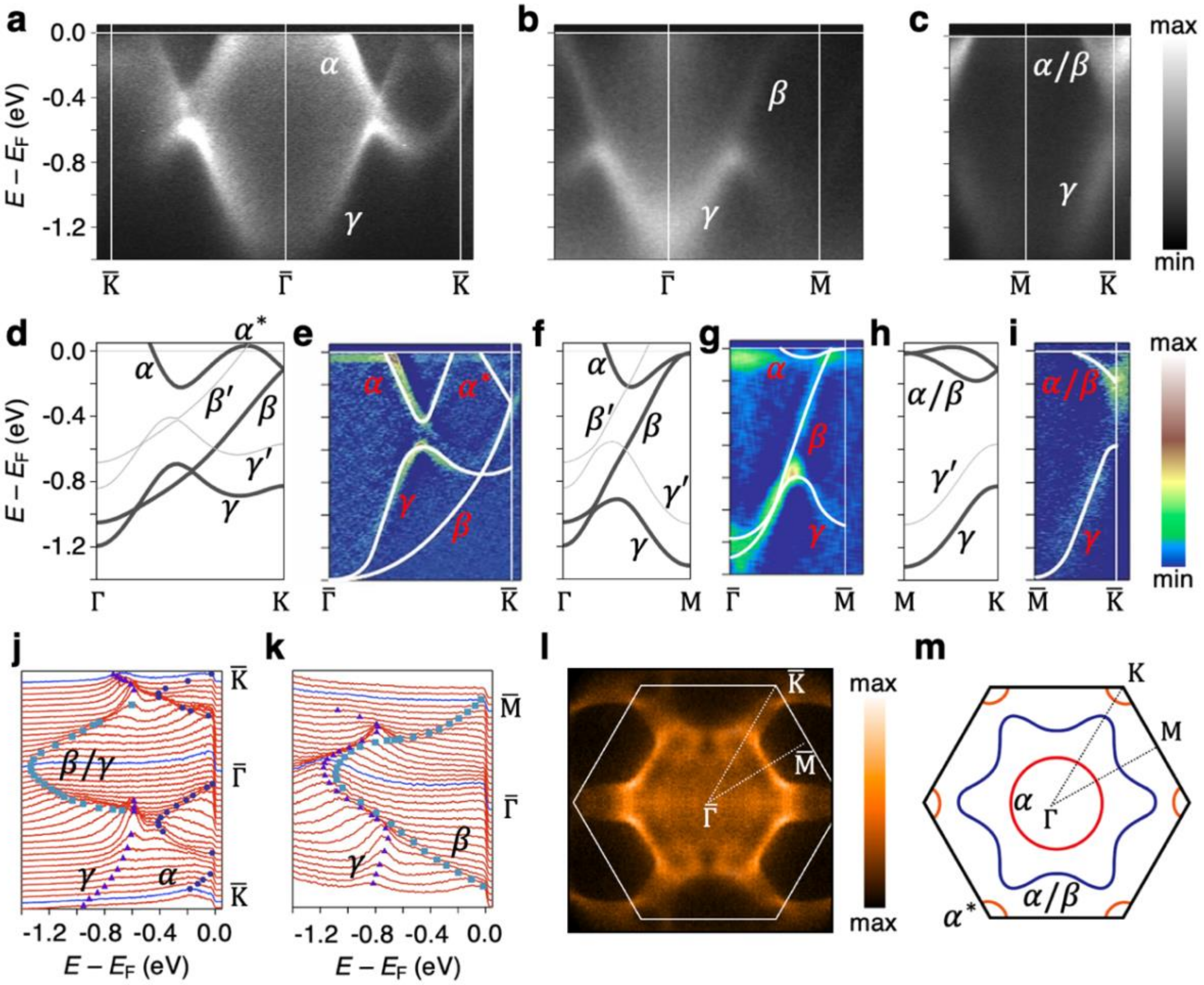

**Fig. 2 Electronic structure of LaCl revealed by ARPES. a–c.** ARPES spectra measured with 75 eV photon energy along $\bar{\Gamma}-\bar{K}$ (a), $\bar{\Gamma}-\bar{M}$ (b), and $\bar{M}-\bar{K}$ (c), revealing three reconstructed dispersive band groups labelled $\alpha$, $\beta$, and $\gamma$, respectively. The $\alpha$ and $\beta$ bands become nearly degenerate along $\bar{M}$–$\bar{K}$. **d, f, h.** Spin-polarized DFT band structures along $\Gamma$–$K$ (d), $\Gamma$–$M$ (f), and $M$–$K$ (h), with primed labels denoting spin-split branches and $\alpha^*$ indicating an additional $\alpha$-derived branch near the zone boundary. **e, g, i.** Second-derivative ARPES maps corresponding to $\bar{\Gamma}$–$\bar{K}$ (e), $\bar{\Gamma}$–$\bar{M}$ (g), and $\bar{M}$–$\bar{K}$ (i); white curves are guide-to-the-eye lines tracing the measured dispersions. **j, k.** Stacked energy distribution curves (EDCs) along $\bar{\Gamma}-\bar{K}$ (j) and $\bar{\Gamma}-\bar{M}$ (k), respectively, with peak positions overlaid to track the band dispersions. The dark-blue circles correspond to the $\alpha$ band, the blue triangles correspond to the $\gamma$ band and the light-blue squares correspond to the $\beta$ band. **l.** Fermi-surface map of LaCl, with the Brillouin zone outlined by the white hexagon. **m.** Schematic drawing of the Fermi surface, showing a $\bar{\Gamma}$-centered $\alpha$ hole pocket (red), a warped $\beta$-derived electron pocket (blue), and additional $\alpha^*$-derived pockets near the zone

boundary (orange). ARPES data shown in this figure were taken using 75 eV photons at temperature $T = 6$ K (panels **a,c,l**) and $T = 7$ K (panel **b**).

**Hopping pathways formed between the La sites and the IAEs**

The three-site dice-lattice model developed for YCl, in which the standalone AEL solely defines the effective lattice, fails to reproduce the dispersive bands of LaCl observed by ARPES. Orbital-projected DFT calculation of LaCl reveals substantial contributions from the La $d_{x^2-y^2}$ and $d_{xy}$ orbitals below $E_{\mathrm{F}}$ (Fig. S9). Polarization-dependent ARPES supports this enhanced $d$-orbital character (Fig. S11). Furthermore, the maximally localized Wannier functions (MLWFs) confirm the Wannier centers of $m = \pm 2$ orbitals are concentrated on the La cations, whereas the $m = 0$ orbitals are still preserved on the IAE sites (Fig. 3e and g, and see Supplementary Information Fig. S12 and Table S8 for further information). A satisfactory description of LaCl therefore requires a five-site basis comprising the three IAE sites A, B and C and two La-centered sites D and E. In this basis, the Hamiltonian can be written in block form as:

$$H_0(\boldsymbol{k}) = \begin{pmatrix} h_{\mathrm{IAE}}(\boldsymbol{k}) & V_{\mathrm{IAE-atom}}(\boldsymbol{k}) \\ V^{\dagger}_{\mathrm{IAE-atom}}(\boldsymbol{k}) & h_{\mathrm{atom}}(\boldsymbol{k}) \end{pmatrix}, \tag{1}$$

where $h_{\mathrm{IAE}}(\boldsymbol{k})$ describes the dice-lattice IAEs network, $h_{\mathrm{atom}}(\boldsymbol{k})$ describes the honeycomb-like La-centered states, and $V_{\mathrm{IAE-atom}}(\boldsymbol{k})$ contains the hopping processes that entwine the two subsystems. The resulting five-site model captures the principal dispersive manifolds observed by ARPES (Fig, 3a,f).

The decisive difference from YCl lies in the off-diagonal coupling $V_{\mathrm{IAE-atom}}(\boldsymbol{k})$. The dominant hopping processes involve not only channels between the IAE sites (A-C, B-C), but also direct IAE-La pathways (A-D, A-E), indicated by the red arrows in Fig. 3g. As a result, the effective lattice connectivity is fundamentally rewired. In the dice-lattice regime, as realized in YCl, the rim sites (A/B sites) only connect with the hub site (C sites), and the flat band originates from isolated antibonding states localized on the rim sites [60] . In LaCl, however, additional hopping pathways are formed between the rim sites and the La sites, and thereby fully reconfigure the lattice network. The outcome is the highly dispersive low-energy manifold as shown in Fig. 3a,f and the reconstructed band topology as shown in Fig. 1d,g. For YCl, the Berry curvature is concentrated at K valley, where each gapped Dirac crossing contributes ($\pm\frac{1}{2}$) to the Chern number. Additional band crossings along the K-Γ path give Berry-curvature contributions with the same sign, leading to a total Chern number of $|C| = 4$). By contrast, in LaCl, the dominant band inversion occurs along the $\Gamma$-M path. The Berry curvature generated around all anticrossing regions has the same sign, as shown in Fig. 1g, and each gapped Dirac point contributes $(\pm\frac{1}{2})$ to the Chern number, giving rise to $|C| = 3$).

Another signature of the reconfigured lattice network is the emergence of a saddle-point dispersion near the $\overline{\mathrm{M}}$ point. ARPES resolves the band approaching this feature near $E_{\mathrm{F}}$ (Fig. 3c,d), while the five-site model proves this van Hove singularity (vHs) (Fig. 3f). This vHS replaces the flat-band feature found in YCl and becomes a key band feature in the La–IAE hybridized hopping network. A $\boldsymbol{k} \cdot \boldsymbol{p}$ model at M point further confirms this feature (see Supplementary Information S7).

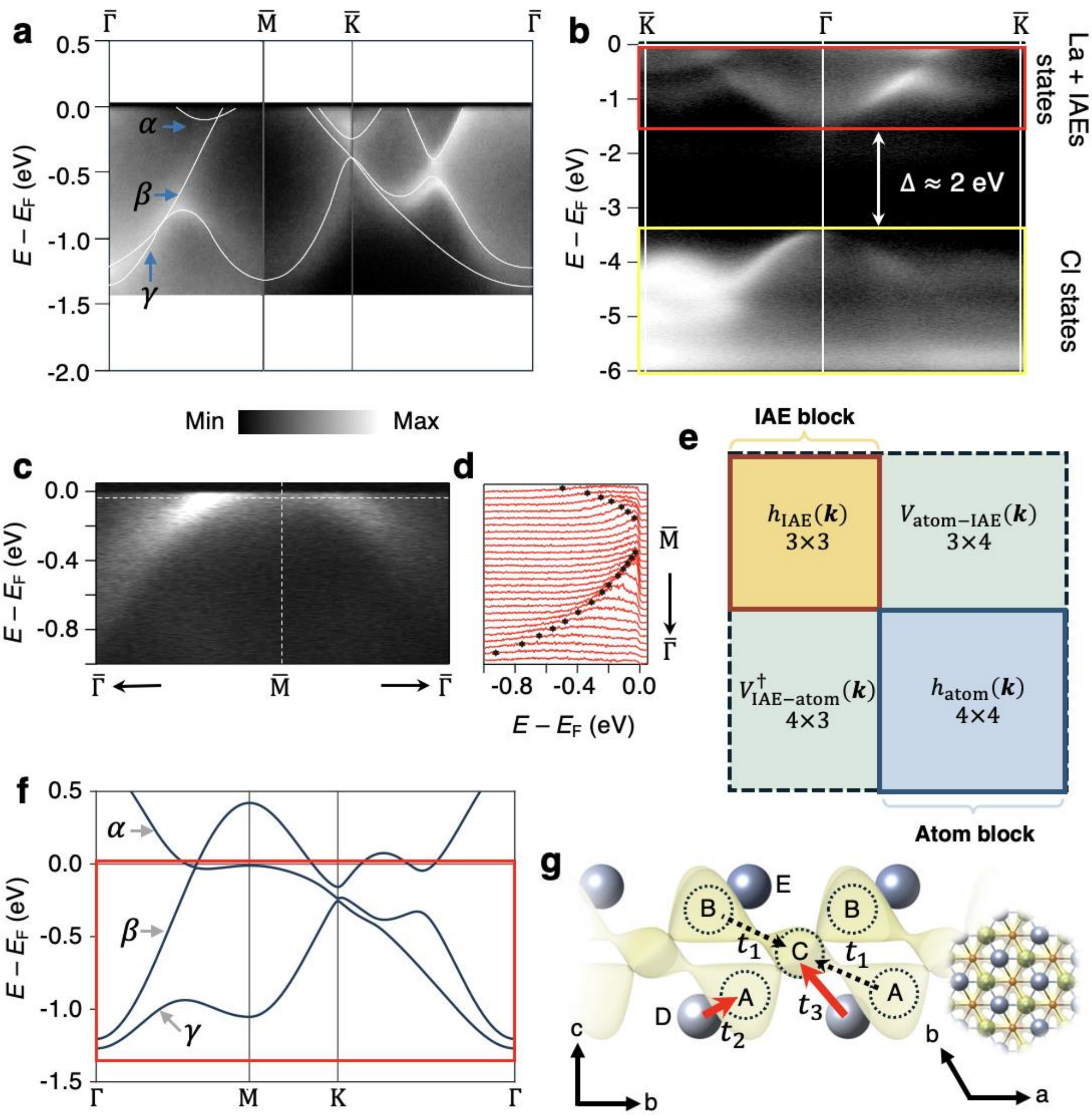

**Fig. 3 The entwined lattice with strong coupling between the anionic electron sublattice and cations in LaCl. a.** Composite ARPES spectrum of LaCl assembled from high-symmetry cuts. The white-dashed curves are the guide-to-the-eye lines for the dispersions. **b.** Wide-energy ARPES spectrum highlighting the separation between the low-energy La/IAE-derived states (red box) and the Cl-derived valence bands (yellow box), which are well separated by an energy gap of $\Delta \approx 2$ eV. **c**. Zoom-in ARPES spectrum along the $\bar{\Gamma}$–$\bar{\mathrm{M}}$ direction, revealing the enhanced spectral intensity associated with the van Hove singularity (vHS) near the $\bar{M}$ point. **d**. Stacked EDCs extracted from the region shown in **c** with peak positions tracking the band dispersion and confirming the band evolution near the vHS. **e.** Schematics of Hamiltonian of LaCl. The Hamiltonian is of block matrix

format, in which the AEL (yellow block) and atomic framework (blue block) are directly coupled. **f.** TB band structure of the minimal model for LaCl. The red box indicates the energy window probed by ARPES. A van Hove singularity shows up in the TB band structure at the M point. **g.** Schematic of the effective tripartite/five-site (A/B, C, D/E) Hamiltonian. Sites A-C represent the anionic electron lattice, while D and E are the La cation sites. Red arrows indicate the new La–IAE hopping pathways introduced in LaCl. The black dashed arrows indicate the original hoppings between the IAE sites (A-C, B-C).

**Microscopic origin of the entwined lattice**

The appearance of entwined lattice in La-based compounds can be qualitatively explained by the anomalous size of the interstitial void that accommodates the IAEs. As shown in Fig. 4a, across the REX family, the intralayer RE–RE distance increases monotonically with RE ionic radius, consistent with a conventional structural trend (see also Supplementary Information Fig. S3 and Table S6). By contrast, the interlayer RE–RE separation displays a clear anomaly in the La-based compounds. Instead of continuing to increase with cation size, it saturates so that the interlayer distances remain nearly identical to those of the Y-based compounds. Such behavior leads to a redistribution of the electron density, which is further evident in the ELF comparisons between YCl and LaCl (Fig. 4c,d), in which the Y cations achieve closed-shell configurations, with minimal coupling to the adjacent IAE sites, whereas the La cations display more anisotropic distributions and enhanced hybridization with both sites A and B IAEs.

These regimes are reflected by the schematics in Fig. 4b. When the cation states are energetically and spatially separated from the IAEs, the low-energy hopping network is defined predominantly by the interstitial sites, producing a standalone AEL, as realized in the Y- and Sc-based compounds. As the out-of-plane cation separation approaches its lower structural limit, enhanced cation–IAE hybridization activates additional hopping pathways that extend beyond the AEL. The interstitial-electron and atom-centered states then form a single entwined electronic lattice, as realized in LaCl. As a direct manifest of this scheme, a numerical interpolation from the standalone AEL to entwined lattices shows the gradual shift from dice-lattice electronic structure of YCl to reconstructed dispersive bands of LaCl with increasing shift in geometry phase factors associated with $d_{x^2-y^2}/d_{xy}$ orbitals (Supplementary Information Fig. S16). CdCl provides a contrasting atomic-lattice-dominated limit, in which its low-energy electronic states are concentrated predominantly around the Cd sites, and a distinct interstitial electron lattice is no longer evident. Consequently, the characteristic Dirac-cone dispersion appears following the honeycomb lattice geometry (Fig. 4e). Therefore, the REX-related layered compounds present a general scheme for IAE-induced effective lattice, evolving from a standalone AEL, through an entwined anionic–cationic lattice, and ultimately toward the conventional atomic lattice that governed electronic structures (Fig. 4c-e).

This perspective places LaCl in a broader context of electronic-structure engineering. Twisted bilayer graphene, for example, became influential not only because it produces flat bands [27, 28], but because it established moiré geometry as a powerful design principle for reshaping electronic structures [55-58]. In an analogous spirit, LaCl shows that the AEL in an electride is not merely a fixed structural motif or a material-specific feature. Rather, it can act as a reconfigurable lattice-like degree of freedom that reshapes the low-energy electronic structure without changing the crystallographic symmetry or electron filling. The identification of LaCl therefore reveals a general principle: anionic-electron lattices can be actively reorganized through their coupling to the atomic framework, providing an intrinsic crystalline route to band-structure engineering.

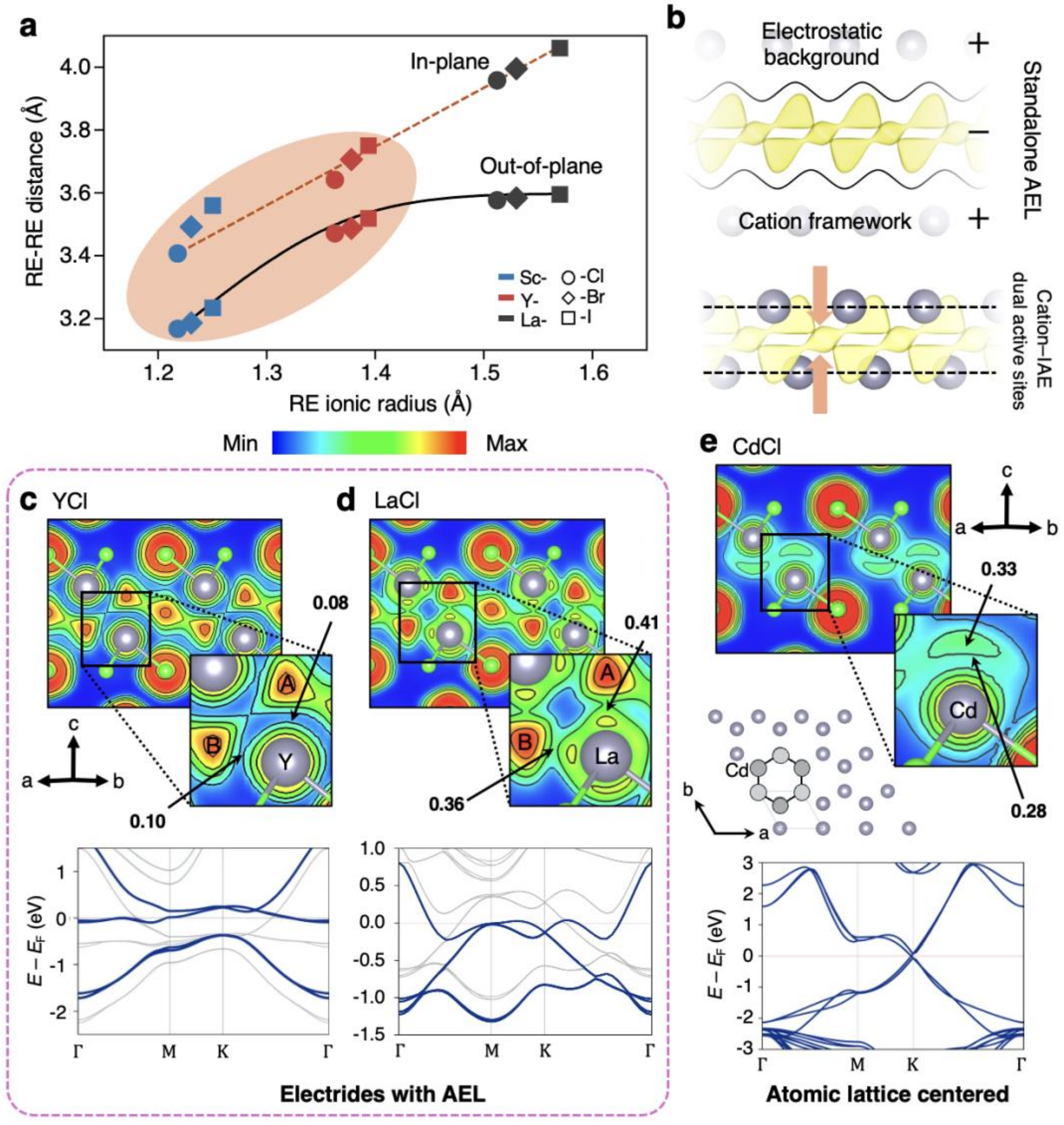

**Fig. 4 Structural trend for the emergence of entwined lattices in the REX electride family. a.** Survey of the REX electride family showing RE–RE distance as a function of cation ionic radius. The in-plane RE–RE separation increases monotonically with ionic radius (dashed orange line), whereas the out-of-plane separation (solid black line) exhibits weak scaling and deviates from the hard-sphere trend in the La-based compounds, delineating a crossover from the standalone AEL regime (shaded area) to the entwined-lattice regime (La-based compounds). **b.** Conceptual schematic plot for the relationship between atomic framework, standalone AELs, and entwined lattices. In the standalone AEL (top), excess electrons occupy interstitial regions and form an electron lattice while the ionic slabs act primarily as an electrostatic background. In the entwined lattice exemplified by LaCl (down), cation sites become active, enabling direct coupling between the AEL and the atomic lattice. **c-e.** Evolution from standalone AEL to atomic-lattice dominated electronic structure. **c.** In YCl, the excess electrons form a standalone AEL with dice-lattice connectivity. The atomic framework provides an electrostatic background that confines the IAEs into localized interstitial sites with discrete ELF maxima, but the cation states remain essentially closed shell and do not directly participate in the low-energy hopping network. **d.** In LaCl, the reduced interlayer spacing enhances cation–IAE coupling. The La-derived states become active in the low-energy electronic structure and provide additional hopping channels for the IAEs, producing an entwined lattice of anionic electrons and cationic states. **e.** In the atomic-lattice centered condition, exemplified by CdCl, the excess charge is distributed more evenly around the cations rather than being localized at well-defined interstitial sites. Accordingly, the low-energy electronic structure is governed mainly by the atomic lattice, giving rise to the regular honeycomb-lattice bands associated with the cation lattice structure, shown in the middle-left inset.

## CONCLUSION

By combining ARPES, DFT calculations, and TB modeling, we show that the AEL in layered electride can provide a novel tuning knob for electronic structure. This is exemplified by the isostructural layered electrides YCl and LaCl, in which YCl hosts a standalone AEL that produces dice bands, whereas LaCl exhibits a reconstructed, dispersive electronic structure with a vHS near $E_F$. The LaCl electronic structure cannot be captured by an IAE-only model and instead requires explicit hybridization between the AEL and the La cation sites, which rewires the effective connectivity and fully reshapes the electronic structure. These results establish the entwined lattice of atoms and anionic electrons as a route to reconfigure lattice topology without changing crystallographic symmetry and electron fillings. Across the REX family, saturated interlayer separation and enhanced low-energy cation orbital participation correlate with the crossover from standalone AEL to entwined lattices, providing practical criteria for engineering emergent electronic lattices.

**Acknowledgements**

The work conducted at the Hong Kong University of Science and Technology (Guangzhou) was supported by NSFC-Young Scientists Fund (No. 12304093, No. 12447158), Guangdong Natural Science Fund-General Program (No. 2025A1515010667), and Start-up Fund of HKUST(GZ) through Grant No. G0101000127 and No. G0101000263. The Modern Matter Laboratory (MML), Green e Materials Laboratory (GeM) and the Materials Characterization and Preparation Facility (MCPF) at HKUST(GZ) provided necessary instruments for the crystal synthesis and characterizations. We acknowledge the support provided by the Bloch beamline at MAX IV Laboratory and beamline BL03U at Shanghai Synchrotron Radiation Facility.

**Author contributions**

S.G. and H.Li. conceived the project. S.G. and H.Li. led the ARPES measurement and analysis. X.W., R.G., Q.W., F.C., K.L., K.A., M.Y., Z.L.,S.H. and T.B. helped with the ARPES measurement. X.W. performed the single crystal synthesis and the crystal characterizations. S.G. carried out the DFT calculations. J.Z. and B.T.Z. constructed the TB modeling. C.Q. T.Y. and H. Liang helped to give suggestions. S.G. and H.Li. did the majority of the paper writing, with contributions from all coauthors. H.Li. directed the overall project.

**Competing interests**

The authors declare no competing interests.

**Data and materials availability**

The data that support the findings of this study are available from the corresponding authors on reasonable request.